\documentclass[aps,prb,reprint]{revtex4-2}

\usepackage[pdftex]{graphicx}
\usepackage{dcolumn}
\usepackage{bm}
\usepackage[english]{babel}
\usepackage{amssymb,amsmath}
\usepackage{mathtools}
\usepackage{physics}
\usepackage{float}
\usepackage{longtable} 
\usepackage[caption=false]{subfig}
\usepackage{hyperref}

\begin{document}

\preprint{APS/123-QED}

\title{Predicting the Curie temperature of magnetic materials \\ with automated calculations  across chemistries and structures}
\author{Marian Arale Br\"annvall}
\author{Gabriel Persson}
\author{Luis Casillas-Trujillo}%
\author{Rickard Armiento}%
\author{Bj\"orn Alling}%
\affiliation{
 Department of Physics, Chemistry, and Biology (IFM), Link\"oping University, 581 83 Link\"oping, Sweden}%

\date{\today}

\begin{abstract}
We develop a technique for predicting the Curie temperature of magnetic materials using density functional theory calculations suitable to include in high-throughput frameworks. We apply four different models, including physically relevant observables and assess numerical constants by studying 32 ferro- and ferrimagnets. With the best-performing model, the Curie temperature can be predicted with a mean absolute error of approximately 126 K. As predictive factors, the models consider either the energy differences between the magnetic ground state and a magnetically disordered paramagnetic state, or the average constraining fields acting on magnetic moments in a disordered local moments calculation. Additionally, the energy differences are refined by incorporating the magnetic entropy of the paramagnetic state and the number of nearest magnetic neighbors of the magnetic atoms. The most advanced model is found to extend well into Fe$_{1-x}$Co$_x$ alloys, indicating the potential efficacy of utilizing our model in designing materials with tailored Curie temperatures by altering alloy compositions. This examination can illuminate the factors influencing magnetic transition temperatures in magnetic materials and provide insights into how they can be employed to make quantitative predictions of Curie temperatures. Our approach is not restricted to specific crystal structures or chemical compositions. It offers a more cost-effective alternative, in terms of human time and need for hands-on oversight, to other density functional theory methods for predicting the Curie temperature. As a result, it provides a practical strategy for conducting high-throughput screening for new technologically applicable magnetic materials. Alternatively, it can complement machine-learning-based screening of magnetic materials by integrating physical principles into such approaches, thereby enhancing their prediction accuracy.
\end{abstract}

\maketitle


\section{\label{sec:intro}Introduction}
For a magnet to be of practical use, it must maintain its magnetic properties at operational conditions. For ferro- and ferrimagnetic materials, the temperature at which they transition into a  macroscopically non-magnetic state, due to a magnetic order-disorder transition, is called the Curie temperature ($T_\text{C}$). The applicability of the magnet is therefore dependent on $T_\text{C}$ and the optimal value will depend on the application. For magnetic sensors in aeroengines or permanent magnets in electric motors, for example, $T_\text{C}$ must significantly exceed room temperature \cite{rokicki_high_2021,podmiljsak_future_2024}. Whereas, in the case of magnetic refrigeration, it is preferable to have a magnetic transition occurring near room temperature \cite{brown_magnetic_1976,gschneidner_thirty_2008}. When searching for new magnetic materials for technological use, it is therefore crucial to be able to predict $T_\text{C}$. As Nelson and Sanvito point out, most magnetic compounds have a  $T_\text{C}$ below room temperature. Hence, merely searching for new magnets without information about $T_\text{C}$ runs the risk of mostly finding candidates that are not very useful \cite{nelson_predicting_2019}. 

There are both experimental and theoretical approaches to determine $T_\text{C}$, each with their own limitations when considering their applicability in the exploration of a diverse range of materials. To experimentally investigate a magnetic material, the alloy must first be synthesized, and then one can study the magnetic properties using different measurement methods, e.g., using a vibrating-sample magnetometer, a superconducting quantum interference device, or a magnetic torsion balance \cite{fabian_measuring_2013, gramm_squid_1976}. This procedure  is time consuming, costly, and not feasible to employ for investigating a wide range of materials across diverse chemistries. Theoretical investigations of $T_\text{C}$ typically involve constructing a Heisenberg Hamiltonian by deriving the exchange interactions between magnetic moments using the method developed by Liechtenstein et al. \cite{liechtenstein_local_1987, ruban_atomic_2004, drautz_spin-cluster_2004, lindmaa_exchange_2013}. The Curie temperature can then be obtained through, e.g., Monte Carlo simulations or a mean-field approximation 
\cite{liechtenstein_local_1987,halilov_adiabatic_1998, lezaic_first-principles_2007,alling_effect_2009,alling_theory_2010,thore_magnetic_2016,vishina_high-throughput_2020,vieira_high-throughput_2021}. The Liechtenstein method of extracting the exchange interactions is cumbersome to apply across a diverse set of complex structures in a high-throughput manner. Furthermore, the need for considerable human hands-on decisions regarding aspects such as magnetic reference state, number of included coordination shells, and effects of local lattice relaxations, has limited studies to treatments of a single or a few systems per study \cite{ halilov_magnon_1997, kubler_ab_2006, rusz_exchange_2006, kormann_free_2008}.
In particular, the current methodology for determining $T_\text{C}$ faces challenges in accurately predicting values for compounds characterized by strongly correlated 4f electrons \cite{turek_ab_2003,miyake_quantum_2018, miyake_understanding_2021, ghosh_unraveling_2023}. Resolving this challenge often necessitates the adoption of methods such as the integration of density functional theory (DFT) with dynamical mean-field theory and/or the DFT+U method \cite{kvashnin_exchange_2015,locht_standard_2016,delange_crystal-field_2017,szilva_quantitative_2023}. When navigating through wide-ranging materials, this introduces an additional layer of complexity.

To overcome these difficulties, different machine-learning (ML) approaches have been developed to predict the critical temperature \cite{dam_important_2018,zhai_accelerated_2018,nguyen_regression-based_2019,nelson_predicting_2019,kabiraj_high-throughput_2020, hu_searching_2020,long_accelerating_2021,lu_--fly_2022,singh_physics-informed_2023,belot_machine_2023, liu_prediction_2024}. The big advantage of these types of methods is that the predictions are very cheap and more computationally heavy calculations or experimental work can be done based on these predictions. However, ML models often operate as black boxes, making it challenging to understand the underlying mechanisms driving the predictions \cite{liu_materials_2017}. Moreover, the applicability of ML models for predicting critical temperatures remains to be proven, especially for, e.g., disordered systems and the effects of alloying. Nelson and Sanvito developed ML models for predicting $T_\text{C}$ using a few different ML algorithms  and the experimental data of about 2500 known ferromagnets \cite{nelson_predicting_2019}.  They developed a model which predicts $T_\text{C}$  with an accuracy of approximately 50 K. They also evaluated the ability of the model to predict $T_\text{C}$ of some binary alloys and the ternary Al-Co-Fe system with not fully convincing results. In a study by Belot et al., they developed machine-learning models using a random forest and a k-NN model \cite{belot_machine_2023}. They found that for binary alloys, the ML model did not manage to interpolate smoothly between points with known experimental $T_\text{C}$, instead, it had a tendency to drift towards the average $T_\text{C}$ of the training data. To reduce this issue they trained a new model only on materials with $T_\text{C}$ greater than $ 600$ K, hence raising the average $T_\text{C}$. In both these studies, the models were only based on chemical composition. Attempts were made to incorporate structural data, giving an increased number of features at the cost of less available training data. The result was a model which could not outperform the chemical-composition-only models \cite{nelson_predicting_2019, belot_machine_2023}. The aim of this work is to investigate a different approach of predicting $T_\text{C}$; a method operating at the level of DFT, but without the need of a detailed modeling of the magnetic interactions. This method can be used both as an initial screening to identify candidate materials to investigate with more expensive in-depth calculations, or as a means to incorporate physics-based insights into a machine-learning model.

We should make clear what we mean by the critical temperature or, more specifically, the Curie temperature in this work. The critical temperature of a magnetic material is defined as the temperature above which the long-range magnetic ordering vanishes \cite{Ashcroft76}. For ferro- and ferrimagnets, this means that there is no spontaneous bulk magnetization. However, the spontaneous bulk magnetization also vanishes in a transition from the ferromagnetic to antiferromagnetic state, which would be an order-order transition. Our approach gives the magnetic order-disorder transition temperature. Hence, if a material undergoes a transition from the ferromagnetic to the antiferromagnetic to the paramagnetic state, our approach will find the temperature corresponding to the transition to the paramagnetic state, where in fact the relevant transition for applications may be the lower ferromagnetic to antiferromagnetic temperature.

Our approach offers a physics-based way of estimating $T_\text{C}$ which illuminates the contributing factors of the transition temperature. The approach is general, in that it is not constructed around a specific crystal structure or chemical family of materials. We explore four different models that, in one way or another, use  either the energy difference between the magnetic ground state and the magnetically disordered paramagnetic state or the strength of the constraining fields necessary for modeling the paramagnetic state. To model the paramagnetic phase, we employ the disordered local moments model \cite{gyorffy_first-principles_1985, alling_effect_2010}. For the ground-state energy, the magnetic ground state must be determined, which in the general case can be of various types, such as ferro-, ferri-, and antiferromagnetic, or even more exotic noncollinear structures. In the work by Horton et al., a workflow for predicting the collinear magnetic ground state using DFT was developed \cite{horton_high-throughput_2019}. By generating multiple plausible magnetic orderings and ranking them based on symmetry, they investigated several experimentally known magnetic materials with nonferromagnetic ground-state ordering. The corresponding experimental magnetic ordering was found in 60 \% of the cases and was correctly predicted as nonferromagnetic in 95 \% of the cases. Galasso and Oganov developed a similar workflow for determining the collinear ground-state structure of magnetic materials. Additionally, the Hubbard U in DFT+U calculations and the Curie temperature through Monte Carlo simulations can be determined in their workflow \cite{galasso_automag_2023}. Their method was tested on three known collinear antiferromagnets. The work of Ehn et al. \cite{ehn_first-principles_2023} describes a technique for a ground-state search with noncollinear magnetism. In this approach, the magnetic moment directions are initialized as randomly distributed but without any constraints, and the atomic positions as ideal. The relaxations are subsequently performed, allowing for both atomic positions and the directions and magnitudes of the magnetic moments to relax. The magnetic moments will consequently change sizes and rotate toward their preferred configuration. In this work, we use this latter technique for determining the ground-state energy, where we focus on materials with experimentally determined ferro- or ferrimagnetic ground state, however, allowing for DFT to find other ground states. 

Based on mean-field arguments, there is a correlation between the energy difference and $T_\text{C}$ \cite{liechtenstein_magnetic_1983, gyorffy_first-principles_1985,turek_ab_2003,gyorffy_first-principles_1985, bergqvist_theoretical_2007, wasilewski_curie_2018}. The relation between the strength of the constraining fields and $T_\text{C}$ is motivated by their connection to the average magnetic interactions, $J_0$, acting on each magnetic moment. Furthermore, we combine these results with structural information and information about the magnetic entropy in the paramagnetic state to draw a physical picture of the factors that play a role in determining $T_\text{C}$. The insights from these investigations are then used to construct models to predict $T_\text{C}$ of ferro- and ferrimagnetic materials. This approach is a more general way of estimating $T_\text{C}$ compared to the theoretical methods described previously and it provides insights that may get lost in predictions solely made with machine learning methods.

\section{\label{ref:background} Methodological Background}
\subsection{Energy Difference and the Curie Temperature \label{ref:MFA_MC_NN}}
The Curie temperature can be estimated using both the mean-field approximation (MFA) and Monte Carlo (MC) simulations. In  a mean-field treatment of the magnetic system we may start from the semiclassical Heisenberg Hamiltonian,

\begin{equation}
\label{eq:hes_Ham}
    \mathcal{H} = -\sum_{i\neq j} J_{ij} \pmb{e}_i \cdot \pmb{e}_j,
\end{equation}
\noindent
where the sum is over every pair of magnetic moments $i$ and $j$; $\pmb{e}_i$ is the direction of the atomic magnetic moment $i$ and $J_{ij}$ is the pair exchange interaction. In a mean-field treatment, the exchange interactions of one magnetic moment with the whole crystal is examined. The effective exchange parameter is then the sum of all pair exchange interaction parameters between the considered moment and the rest,
\begin{equation}
\label{eq:J_0}
    J_0 = \sum_j J_{0j},
\end{equation}
where the considered magnetic moment is at site 0 \cite{liechtenstein_local_1987}. The energy per magnetic moment can then be expressed in the mean-field framework as
\begin{equation}
\label{eq:E_MFA}
    E^{MFA} = -\frac{2J_0}{M_0^2}\langle\pmb{M}\rangle \cdot \pmb{M},
\end{equation}
where $\pmb{M}$ is the magnetic moment vector and $M_0$ is the magnitude of the considered magnetic moment \cite{liechtenstein_local_1987}. Subsequently, it can be shown that the Curie temperature, $T_\text{C}$, is approximated as
\begin{equation}
\label{eq:T_C_MFA}
    k_\text{B}T_\text{C} = \frac{2}{3}J_0,
\end{equation}
where $k_\text{B}$ is the Boltzmann constant. As a rough estimate, we may then identify the exchange interaction $J_0$ as proportional to the energy difference between an ideal magnetically disordered paramagnetic state and the ferromagnetic state through Equation (\ref{eq:E_MFA}). Combining this insight with Equation (\ref{eq:T_C_MFA}) gives us an estimate of $T_\text{C}$ as \cite{liechtenstein_magnetic_1983,bergqvist_theoretical_2007, turek_ab_2003, wasilewski_curie_2018}

\begin{equation}
\label{T_C_deltaE}
    k_\text{B}T_\text{C} \propto (E_\text{PM} - E_\text{FM}).
\end{equation}
where $E_\text{PM}$ and $E_\text{FM}$ are the energies of the ideal paramagnetic state and the ferromagnetic state, respectively.

Another, more accurate way to estimate the Curie temperature is by means of MC simulations. In this case, the energy can be evaluated based on the Hamiltonian of Equation (\ref{eq:hes_Ham}) and with exchange interactions extracted beforehand from DFT calculations. The $T_\text{C}$ can then be determined, e.g, from the location of the characteristic peak in the temperature dependence of the specific heat at the phase transition.

\subsection{Disordered Local Moment \label{ref:DLM}}
The disordered local moment (DLM) method is an approach for simulating the paramagnetic state in DFT calculations. The method has been implemented both within the coherent potential approximation by Gyorffy et al. \cite{gyorffy_first-principles_1985} and within a supercell approach by Alling et al. \cite{alling_effect_2010}. Within the DLM method, the local magnetic moments are fully disordered, describing an ideal case with no short-range order (SRO). It is assumed that describing the magnetization density as locally collinear in the region around an atom is a valid approximation \cite{sandratskii_noncollinear_1998}. This assumption leads to the atomic moment approximation, where a magnetic moment vector is assigned to each atom, defined as

\begin{equation}
    \mathbf{m}_i = \int_{\Omega_i}\mathbf{m}(\mathbf{r}) d\mathbf{r},
\end{equation}
\noindent
where $\Omega_i$ describes a region around atom $i$ and $\mathbf{m}(\mathbf{r})$ is the magnetization density. As mentioned, in the DLM approach, these atomic magnetic moments are distributed to obtain maximum disorder with minimal SRO and ideally,

\begin{equation}
\label{eq:SRO_DLM}
    \frac{1}{N} \sum_{i,j \in \alpha } \mathbf{e}_i \cdot \mathbf{e}_j = 0, \; \forall \alpha,
\end{equation}
\noindent
where $N$ is the number of magnetic pairs in the summation; $i$ and $j$ denote the atoms in question belonging to a specific coordination shell $\alpha$, and $\mathbf{e}_i$ and $\mathbf{e}_j$ are the unit vectors in the direction of the magnetic moments. 

Two ways of implementing DLM within a supercell approach have been developed to ensure results corresponding to a true ideal paramagnetic state. One may either use the special quasirandom structure (SQS) method \cite{zunger_special_1990} applied to magnetic systems or the magnetic sampling method (MSM) \cite{alling_effect_2010}. In MSM, the magnetic moment directions are randomly distributed (either collinearly or noncollinearly) but not necessarily fulfilling Equation (\ref{eq:SRO_DLM}). Averages of physical properties can then be extracted from multiple of these magnetic samples. These averages are considered to be representative for the disordered state.

\subsection{Constraining Fields \label{ref:constr_fields}}
An arbitrary noncollinear magnetic configuration is not a ground state, therefore, when running noncollinear DFT calculations, the magnetic moments tend to rotate towards the magnetic ground state during the self-consistency cycle. By applying constraining fields to each of the magnetic moments to force them in a certain direction, we can however, simulate the disordered high temperature paramagnetic state.  Ma and Dudarev presented a technique for constraining the directions of the magnetic moments, allowing them to change in size but not sign \cite{ma_constrained_2015}. In this approach, the total energy functional is,

\begin{equation}
\label{eq:tot_energy_constr}
    E = E_0 + E_p = E_0 + \sum_i \lambda_i(\abs{\mathbf{M}_i^F} - \mathbf{e}_i\cdot\mathbf{M}_i^F),
\end{equation}
\noindent
where $\mathbf{M}_i^F = \int_{\Omega_i} \mathbf{m}(\mathbf{r})F_i(\abs{\mathbf{r}-\mathbf{r}_i})d^3r$ and $F_i(\abs{\mathbf{r}-\mathbf{r}_i}) = \sin{(x)}/x$ with $x = \pi\abs{\mathbf{r}-\mathbf{r}_i}/R_i$ to ensure that the magnetization density is smoothly decreased to zero at the atomic sphere boundary. The term $E_0$ is the DFT energy, and $E_p$ is the penalty energy which tends towards zero as $\lambda_i \rightarrow \infty$. The parameter $\lambda$ must be set high enough and often incremented in steps for the energy to be converged. The constraining fields are given by the variation of $E_p$ with respect to the magnetic moment direction \cite{ma_constrained_2015},

\begin{equation}
    \label{eq:constr_fields}
    \pmb{b}_p(\pmb{r}) = -\frac{\delta E_p}{\delta \pmb{m}(\pmb{r})}.
\end{equation}
\noindent
The strength of these constraining fields is connected to the effective exchange parameter, $J_0$, of Equation (\ref{eq:J_0}). This can be understood by considering that the effective magnetic field, in a mean-field treatment, is proportional to $J_0$ and the constraining field strengths are connected to the effective magnetic field \cite{streib_equation_2020}. The magnitude of $J_0$ essentially describes the energy cost of rotating a magnetic moment from its equilibrium direction, thereby influencing how strongly the constraining fields need to act to keep the magnetic moment in this nonequilibrium direction.
\section{Method}
\subsection{Effect of number of nearest magnetic neighbors \label{SRO_NN}}
\begin{figure}
    \centering
    \includegraphics[scale=0.35]{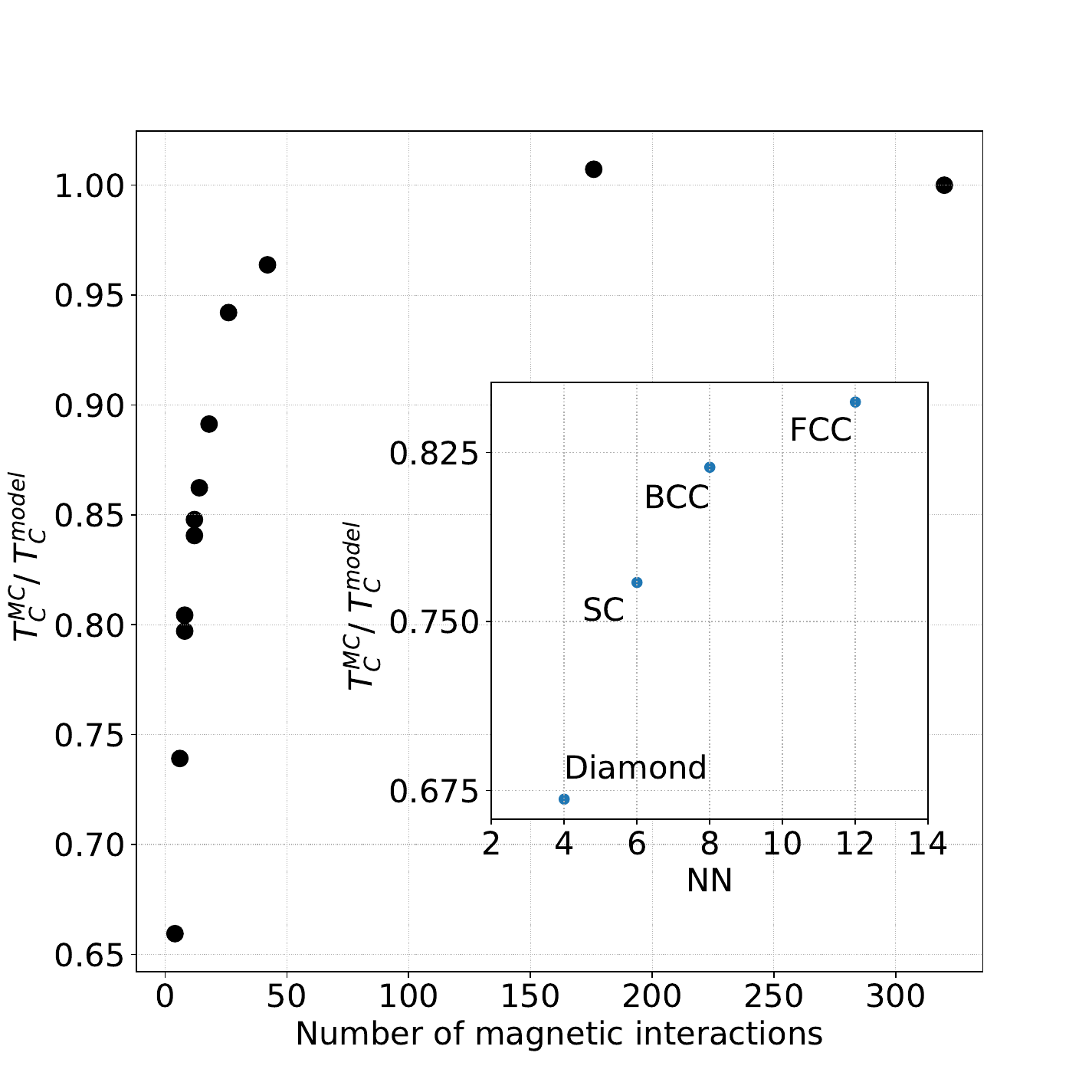}
    \caption{The ratio between the Curie temperature based on specific heat peak from Monte Carlo simulations ($T_\text{C}^\text{MC}$) and the Curie temperature of the simulation with the highest number of magnetic interactions (the mean-field limit, $T_\text{C}^\text{model}$) as a function of the number of interactions. The inset shows the region where diamond, simple-cubic (SC), body-centered-cubic (BCC), and face-centered-cubic (FCC) structures are found. Note that all MC simulations are done for systems having the same $\Delta E$ between the DLM and ground state energies.}
    \label{fig:Heis_MC_NN_ratio}
\end{figure}

The presence of additional (or fewer) magnetic neighbors should have an impact on $T_\text{C}$ through differing short-range-order effects and, consequently, influence the relation between the energy difference, $\Delta E = E_\text{DLM} - E_\text{FM}$, and $T_\text{C}$. This relation can be investigated with MC simulations. By keeping $\Delta E$ fixed and only varying the number of magnetic interactions included in the sum of Equation (\ref{eq:J_0}) we investigate the effect this has on the position of the specific heat peak. Each pair interaction in Equation (\ref{eq:J_0}) is equally strong, independent of the distance between atoms. The number of magnetic interactions is varied from 4 up to 320. The high end of the scale does not represent a realistic number of equally strong magnetic interactions in a real system, but is included to highlight the contrast between a few strong and many weak interactions. Figure \ref{fig:Heis_MC_NN_ratio} shows the ratio between the $T_\text{C}$ from the MC simulations based on the peak of the specific heat curve ($T_\text{C}^\text{MC}$) and $T_\text{C}^\text{model}$ as a function of the number of magnetic interactions.  The temperature $T_\text{C}^\text{model}$ is the Curie temperature of the structure with the highest number of magnetic interactions, since in the limit of infinitely many nearest magnetic neighbors, the result should go towards the mean-field solution.  We clearly see that with fewer nearest neighbors, the discrepancy will be larger between the MC and modeled $T_\text{C}$. 

This observation suggests that one can adjust $T_\text{C}^\text{model}$ based on the number of magnetic interactions to better approximate $T_\text{C}^\text{MC}$. In this work, interpreting the shape of the curve of Figure \ref{fig:Heis_MC_NN_ratio}, we adjust $T_\text{C}^\text{model}$ as 
\begin{equation}
\label{eq:adjustment_NN}
    T_\text{C}^{\text{adj.}} = T_\text{C}^\text{model}\Big(1 - \frac{A}{NN^B} \Big ), 
\end{equation}
where $NN$ is the number of nearest magnetic neighbors of the magnetic atoms, and the parameters $A$ and $B$ are determined by minimizing the difference between adjusted and experimental $T_\text{C}$. This adjustment ensures that the different impact of short-range order effects for different crystal structures and alloy compositions are reflected in the predicted $T_\text{C}$.
\subsection{Disordered Local Moment Calculations \label{DLM_method}}
 For the DLM calculations we employ the supercell approach described in section \ref{ref:DLM}, however, we use neither the traditional SQS- nor the MSM implementation. The number of time-consuming calculations must be kept to a minimum. Since we perform noncollinear magnetic DFT calculations, it is not feasible to conduct calculations for a large number of magnetic samples. For this reason, we generate multiple configurations of randomly directed magnetic moments and choose one where the SRO of the first coordination shell is close to zero (of the order of 0.001 or smaller) for the DFT calculations. The SRO is defined as,

\begin{equation}
    SRO = \Bigg \langle \frac{1}{N_{\text{at}}} \sum_{i=1}^{N_{\text{at}}} \frac{1}{N_{\text{NN}}}\sum_{j=1}^{N_{NN}}\pmb{e}_i\cdot\pmb{e}_j \Bigg\rangle,
\end{equation}
\noindent
 where, $N_{\text{at}}$ is the total number of atoms of the cell and $N_{\text{NN}}$ is the number of magnetic neighbors in the first coordination shell.

 From the DLM calculations, both the constraining fields and the magnitudes of the magnetic moment are collected. The magnetic entropy of the magnetically disordered state is approximated as,

 \begin{equation}
\label{eq:S_mag}
    S^\text{mag} = k_\text{B}\Bigg[\sum_{i=1}^{N_\text{con}}\ln{(m_i + 1)}\Bigg]/N_\text{con},
\end{equation}
\noindent
where $k_\text{B}$ is the Boltzmann constant, $N_\text{con}$ is the number of atoms with constrained magnetic moments, and $m_i$ is the magnetic moment magnitude of the $i$th constrained magnetic moment.
\subsection{\label{ref:exp_Tc}Experimental Curie Temperatures}
The experimental Curie temperatures used in this work are presented in Table \ref{tab:dataset}, with references to the corresponding studies. One should remember that for many systems there are multiple reported Curie temperatures, which signals the sensitivity of the exact value of $T_\text{C}$ on the exact composition or presence of defects. Hence, choosing which value to use when developing predictive models is not obvious. Both Ref. \cite{nelson_predicting_2019} and \cite{belot_machine_2023} used the median Curie temperature when selecting between multiple values in their machine learning model constructions. This would ensure that the value used is an actual measured value, rather than the mean value. In our work, we have chosen one temperature, and for most cases, the differences found are not more than a few kelvin. In all cases the different temperatures are well within the mean absolute errors of our developed models. 
\subsection{\label{ref:workflow} Workflow}
\begin{figure*}
\includegraphics[scale=0.5]{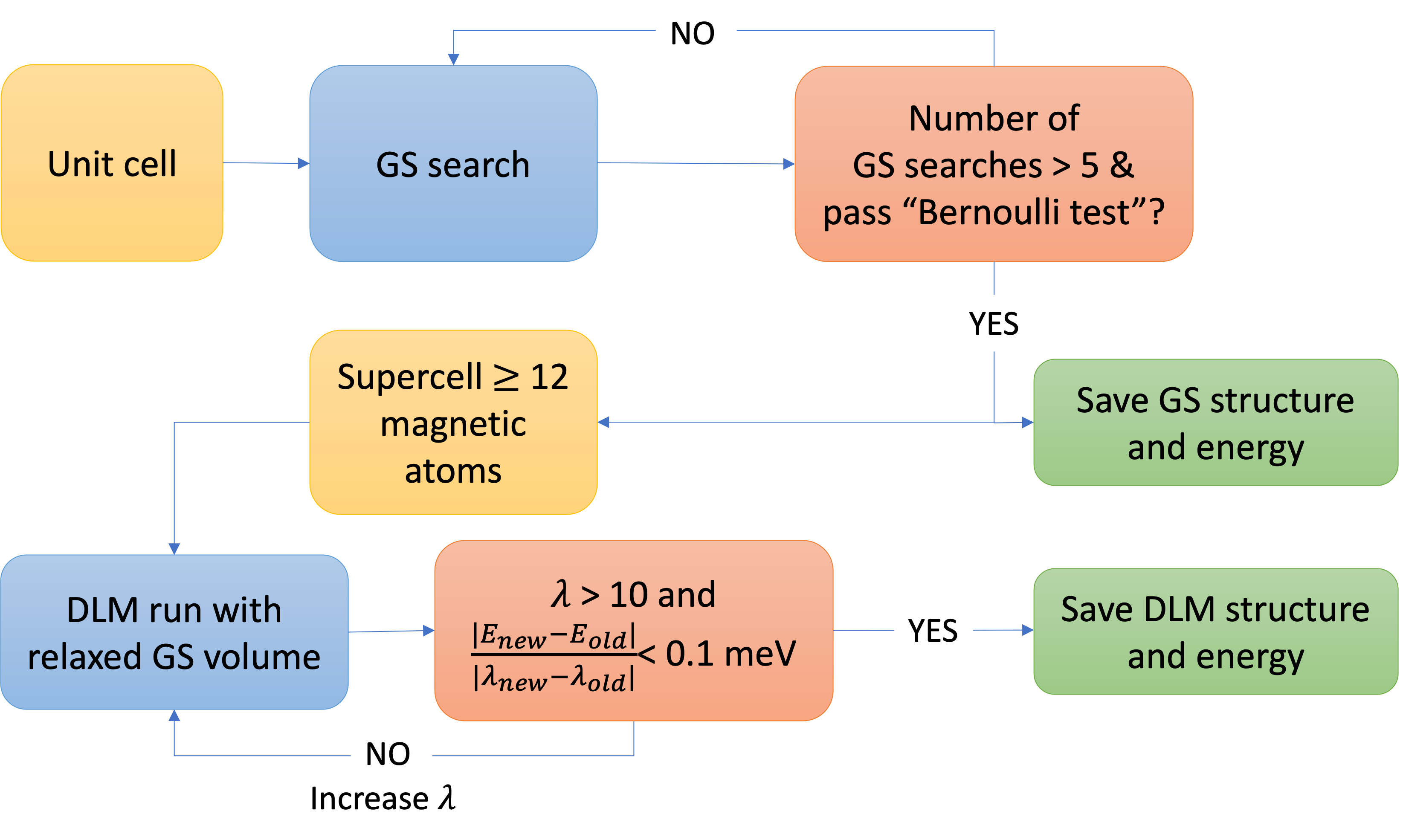}
    \caption{\label{fig:workflow}The workflow of collecting the energy difference and constraining fields of ferri- and ferromagnetic systems. ``Bernoulli test" refers to the check made of the distribution of ground-state (GS) energies as described in the text.}
\end{figure*}

Figure \ref{fig:workflow} illustrates the workflow implemented to collect the energy difference between the ground state  and the paramagnetic state described within the DLM approach. We also obtain the constraining fields from the DLM simulations. 

A system to be investigated, from e.g. the Materials Project \cite{jain_commentary_2013}, is fed into the workflow. If the original structure contains less than 12 magnetic atoms, a supercell is constructed by multiplying the cell in $x$, $y$, and $z$ direction until the supercell contains enough magnetic atoms. The supercell is necessary when simulating the magnetically disordered state using the DLM method. The first ground-state search is subsequently conducted on the original unit cell fed into the workflow. The ground-state search is done using the technique described by Ehn et al. where the directions of the magnetic moments are initially set randomly and are allowed to rotate to their preferred  directions during the structural relaxation \cite{ehn_first-principles_2023}. The choice of unit cell in the ground-state search plays an important role in determining the outcome of the search. A unit cell with only one magnetic atom will, of course, always lead to a ferromagnetic ordering. However, a unit cell with multiple magnetic atoms allows for the exploration of diverse magnetic configurations, such as antiferromagnetic, ferrimagnetic, or complex noncollinear ones.

The first step is to perform five ground-state searches. Based on these, the lowest energy is identified, and the other energies are expressed relative to this minimum. The distribution of the results within 0.025 eV/magnetic atom from the minimum energy determines whether additional ground-state searches should be conducted. If the probability that at least one of the ground-state searches (excluding the minimum) is within 0.025 eV/atom from the minimum value is 90 \% or more, the ground-state search is considered to be finished and the magnetic structure with the lowest energy taken as the ground-state structure. The probability is determined based on the Bernoulli trial framework and the binomial probability function, where a success is an energy within 0.025 eV/magnetic atom from the minimum \cite{boas_mathematical_methods}. The number of ground-state searches is limited to a  maximum of 14. The outcome of this procedure is the energy of the ground-state structure. 

The next step is a DLM run performed without structural relaxations and with the directions of the magnetic moments constrained. The constrain is enforced by a starting $\lambda$-value set to $1$ (see Equation (\ref{eq:tot_energy_constr})). The directions of the magnetic moments are set randomly as described in section \ref{DLM_method}. The constraint is only enforced on atom types where the average magnetic moment is larger than 0.75 Bohr magneton in the ground-state run, smaller magnetic moments will therefore be free to rotate to whichever direction they prefer. Hence, very itinerant systems can not be described in the current workflow; this would require a constraint on the magnitudes of the magnetic moments in the DLM step. The DLM runs continue with the parameter $\lambda$ set to 5, and then 10. Subsequent $\lambda$-steps use the wave functions and charge density of the previous step. Beyond $\lambda = 10$, a check is made of the energy difference between the different $\lambda$-steps. The convergence with respect to $\lambda$-value is examined by calculating the slope 

\begin{equation}
    \frac{\abs{E_{new} - E_{old}}}{\abs{\lambda_{new} - \lambda_{old}}}. 
\end{equation}
The energy with respect to the $\lambda$-parameter is considered converged if the slope is less than 0.1 meV. If it is not converged, the $\lambda$-parameter is raised in steps of 5 until convergence is reached.  

\subsection{\label{ref:model_devel} Model development}
 Four different models for estimating $T_{\text{C}}$ are investigated, denoted as models A, B, C, and D. In model A, the energy difference between the paramagnetic state and the magnetic ground state is used on its own; model B uses the average strength of the constraining fields necessary for modeling the paramagnetic state; model C uses the energy difference divided by the magnetic entropy of the paramagnetic state  as given by Equation (\ref{eq:S_mag}); and model D adjusts the energy difference divided by the magnetic entropy according to Equation (\ref{eq:adjustment_NN}) to incorporate the effect of the number of strong magnetic interactions, and thus the relative effect of short-range order. In each case, a linear model is fitted between the DFT output data and the experimental $T_\text{C}$. The linear regression is implemented using Scikit-Learn \cite{scikit-learn}. For model D, an additional step is needed to incorporate the effect of the number of strong magnetic interactions. The energy difference divided by the magnetic entropy (corresponding to $T_\text{C}^\text{model}$) is adjusted according to Equation (\ref{eq:adjustment_NN}). The parameters A and B are optimized using the Nelder-Mead method \cite{nelder_simplex_1965} by minimizing the mean absolute error of the adjusted model compared to the experimental Curie temperature, $ \langle \abs{T_C^{\text{adj.}}- T_{C}^{\text{exp.}}}\rangle$. 
\section{Computational details}

To obtain the energies, magnetic moments, and magnetic constraining fields, we use DFT calculations as implemented in the Vienna Ab initio Simulation Package \cite{kresse_ab_1993, kresse_ab_1994, kresse_efficiency_1996, kresse_efficient_1996} using  projector augmented-wave potentials (PAW) \cite{blochl_projector_1994,kresse_ultrasoft_1999} and the Perdew-Burke-Ernzerhof (PBE) generalized gradient approximation (GGA) for approximating the exchange-correlation functional \cite{perdew_generalized_1996}. The energy cutoff is set to 450 eV and a Monkhorst-Pack k-point grid with a density of 20 k-points/\r{A} is used \cite{monkhorst_special_1976}. The ground-state search as suggested in Ref. \cite{ehn_first-principles_2023} is conducted on the unit cells obtained from the Materials Project \cite{jain_commentary_2013}. 

In the ground-state search, the magnetic moments are initially set to 3 Bohr magneton ($\mu_B$) with random, noncollinear, directions. The magnetic moments are then allowed to relax in both magnitude and direction, along with the structural relaxation.

To simulate the paramagnetic state, the DLM approach is used. A supercell of at least 12 magnetic atoms is constructed. The calculations are spin-polarized, utilizing noncollinear magnetism. The magnitudes are allowed to relax in the DLM simulations, but are initially set to 110 \% of the relaxed average ground-state size of the magnetic moments of the specific atomic species. The directions of the magnetic moments are constrained using the method presented by Ref. \cite{ma_constrained_2015}. 

As a test case, we examine the binary Fe$_{1-x}$Co$_x$ alloys with different Co contents to investigate how well our models estimate their Curie temperatures. For these calculations, the chemical disorder in the bcc and fcc Fe$_{1-x}$Co$_x$ alloys is implemented using the SQS method \cite{zunger_special_1990}. 

\section{Results \label{result}}
 \subsection{Magnetic Ground State Structures}

Table \ref{tab:dataset} presents both the experimentally observed magnetic ground state and the theoretically obtained ground state from the ground-state search for comparison. Most of the ground-state searches reproduce the experimentally found ferro- and ferrimagnetic ground states of the materials considered here. However, some differences can be observed. For hcp Gd, an antiferromagnetic ordering is obtained.  It has been shown that treating the 4f electrons of bulk hcp Gd using the GGA exchange correlation functional gives an incorrect antiferromagnetic ground state instead of the correct ferromagnetic one. One solution for this issue is to use GGA+U to treat the 4f electrons \cite{kurz_magnetism_2002}. However, in this work we have not explored GGA+U to improve this aspect of the DFT results.

Both CrBr$_3$ and CrI$_3$ should be ferromagnetically ordered \cite{jennings_heat_1965, mcguire_coupling_2015, huang_layer-dependent_2017}. These systems are layered and our ground-state searches result in ferromagnetic alignment of the Cr atoms within the layers, however, not between the layers. This is most likely due to weak magnetic interactions between magnetic moments of different layers.

In line with experiments \cite{wei_synthesis_2012, nguyen_fe3o4_2021}, our calculations results in a ferrimagnetic ground state for Fe$_3$O$_4$ where 1/3 of the Fe magnetic moments are antiferromagnetically aligned with the remaining 2/3 (the unit cell contains 24 Fe atoms). The ferrimagnetic structure is not obtained by a simple collinear calculation, which instead results in a ferromagnetic structure with a significantly higher energy. 

Mn$_3$Ge in an L1$_2$-type structure, as investigated here, is reported to have a ferromagnetic ground state \cite{takizawa_high-pressure_2002}, however, our ground-state search do not find this. Instead, we find a magnetic state that can be described as a ferrimagnet or close to an antiferromagnet with total magnetic moment of only 0.90 $\mu_B$ for a unit cell with 3 Mn atoms. A collinear ferromagnetic calculation gives an energy higher than for the structure obtained with the ground-state search. The DFT study of Mn$_3$Ge by Arras et al. \cite{arras_phase_2011} found the ground state to be ferrimagnetic for the L1$_2$ structure, which suggests that some effect is not accounted for in the DFT simulations, but this is not further investigated in this work. 

Cr$_2$CuO$_4$ is reported to be ferrimagnetic \cite{gurgel_magnetization_2013}, and our ground-state search gives this magnetic ordering with a total magnetic moment of 10 $\mu_B$ for a unit cell with 8 Cr atoms. Out of the 8 Cr moments, 6 are ferromagnetically aligned and the remaining 2 are slightly larger and antiferromagnetically aligned with the rest. 

Mn$_4$N is ferrimagnetic \cite{li_fabrication_2008}, and our ground-state search finds a structure similar to the ferrimagnetic structure found in the DFT study by Adhikari et al. \cite{adhikari_first-principles_2018}, with a total magnetic moment of 1.16 $\mu_B$ for the unit cell containing four Mn atoms. However, in our found structure, the face-centered Mn moments are not completely ferromagnetically aligned with each other. 

Fe$_3$BO$_6$ is a weak ferromagnet (or canted antiferromagnet) \cite{wolfe_magnetization_1969, tsymbal_orientation_2006}, and our ground-state search results in a magnetic structure that can be described as ferrimagnetic, resulting in a total magnetic moment of 4.4 $\mu_B$ for a unit cell with 12 Fe atoms, with a lower energy than for the ferromagnetically ordered structure.
\subsection{Assessment of Model Parameters}

\begin{table*}

\caption{\label{tab:dataset} Data for all investigated systems. The reference to where the experimentally data can be found is given next to the system formula. The magnetic ground states are given both from experiment and from our DFT ground-state search where FM = ferromagnetic, FiM = ferrimagnetic, and AFM = antiferromagnetic. The experimental $T_\text{C}$ and the $T_\text{C}$ estimated by model A, B, C, and D. The energy difference between the ground state and disordered paramagnetic state ($\Delta E$) the magnetic entropy ($S^\text{mag}$), and the average constraining field ($B_\text{con}$) are given per constrained atomic magnetic moment.}
\begin{ruledtabular}
\begin{tabular}{ccccccccccc}

 & \multicolumn{2}{c}{Magnetic Ordering } & & \multicolumn{4}{c}{Model $T_\text{C}$ (K)} & \multicolumn{3}{c}{DFT data} \\ \cline{2-3} \cline{5-8} \cline{9-11}

Material & Exp. & DFT & T$_{C}^{\text{exp.}}$ (K) & $T_{\text{C}}^\text{A}$ & $T_\text{C}^\text{B}$ & $T_\text{C}^\text{C}$ & $T_\text{C}^\text{D}$ &$\Delta E$ (meV)& $S^{\text{mag}}$ (meV/K)& B$_{\text{con}}$ (meV) \\
\hline
Co$_3$V$_2$O$_8$ \cite{rogado_kagome-staircase_2002}   & FM\footnote{FM to AFM at ~6 K, AFM to paramagnetic at ~10 K}  & FM& 10 & 425 & 440 & 398 & 366 & 73.8 & 0.108 & 28.7 \\
CrBr$_3$ \cite{jennings_heat_1965}& FM   & FM \footnote{ \label{CrX}FM within layers but not between layers.} & 33 & 139 & 242 & 133 & 132 & 4.6 & 0.118 & 4.5 \\
CrI$_3$ \cite{mcguire_coupling_2015, huang_layer-dependent_2017} & FM & FM \textsuperscript{\ref{CrX}} & 58 & 103 & 189 & 131 & 136 & 8.9 & 0.177 & 0.5 \\
GdN \cite{li_magnetic_1997}& FM & FM & 68 & 211 & 244 & 220 & 197 & 21.4 & 0.120 & 4.9 \\
Cr$_2$CuO$_4$ \cite{gurgel_magnetization_2013}& FiM  & FiM& 122 & 351 & 291 & 313 & 311 & 58.1 & 0.108 & 10.3 \\
GdCd \cite{sekizawa_magnetic_1966}& FM & FM & 262 & 172 & 221 & 164 & 169 & 24.3 & 0.177 & 4.6 \\
GdZn \cite{rouchy_magnetic_1981}& FM  & AFM & 269 & 157 & 237 & 173 & 153 & 8.1 & 0.177 & 4.1 \\
MnP \cite{felcher_magnetic_1966, matsuda_pressure_2016}& FM\footnote{FM from $T >$ 50 K, helical order below 50 K} & FM & 291 & 537 & 803 & 680 & 604 & 101.3 & 0.068 & 72.7 \\
Gd (hcp) \cite{Ashcroft76}& FM  & AFM & 293 & 176 & 222 & 148 & 149 & 14.0 & 0.178 & 2.1 \\
CrO$_2$ \cite{skomski_simple_2008} & FM  & FM & 398 & 508 & 578 & 505 & 449 & 98.1 & 0.092 & 44.9 \\
Mn$_3$Ge (L1$_2$) \cite{takizawa_high-pressure_2002} & FM & AFM/FiM & 400 & 461 & 469 & 435 & 459 & 82.6 & 0.104 & 32.2 \\
Fe$_3$Pt \cite{wallace_magnetism_1971} & FM  & FM& 430 & 505 & 639 & 441 & 476 & 97.4 & 0.110 & 52.2 \\
Fe$_3$C (cementite) \cite{haglund_electronic_1991}& FM  & FM & 488 & 529 & 539 & 642 & 589 & 104.1 & 0.076 & 45.5 \\
Cu$_2$MnIn \cite{oxley_heusler_1963}& FM  & FM  & 500 & 617 & 347 & 473 & 531 & 123.7 & 0.132 & 20.7 \\
Fe$_3$BO$_6$ \cite{wolfe_magnetization_1969} & weak FM & FiM & 508 & 917 & 678 & 669 & 624 & 194.5 & 0.134 & 57.6 \\
MnB \cite{fries_magnetic_2016}& FM  & FM & 579 & 683 & 666 & 771 & 678 & 142.6 & 0.079 & 55.4 \\
FeB \cite{barinov_structure_1991} & FM & FM & 580 & 579 & 691 & 734 & 719 & 115.1 & 0.071 & 65.0 \\
MnSb \cite{takei_magnetic_1963}& FM  & FM  & 600 & 465 & 386 & 377 & 340 & 87.1 & 0.122 & 21.7 \\
Fe$_3$P \cite{gambino_magnetic_1967}& FM  & FM & 716 & 706 & 627 & 810 & 600 & 143.7 & 0.079 & 56.7 \\
Fe$_3$Sn \cite{sales_ferromagnetism_2014}& FM  & FM & 725 & 873 & 763 & 812 & 722 & 183.8 & 0.099 & 67.8 \\
FePd \cite{longworth_temperature_1968}& FM & FM & 729 & 669 & 774 & 552 & 544 & 133.7 & 0.119 & 69.2 \\
Mn$_4$N  \cite{li_fabrication_2008}& FiM & FM & 740 & 450 & 458 & 479 & 538 & 86.3 & 0.091 & 35.0 \\
Co$_3$B \cite{pal_properties_2017}& FM & FM & 750 & 517 & 774 & 949 & 899 & 101.3 & 0.047 & 75.6 \\
Fe$_3$Ge \cite{kanematsu_magnetic_1963}& FM  & FM & 755 & 727 & 732 & 698 & 762 & 153.6 & 0.096 & 63.4 \\
Fe$_4$N \cite{frazer_magnetic_1958} & FM & FM & 761 & 566 & 734 & 564 & 642 & 112.8 & 0.092 & 63.6 \\
FeNi (L$1_0$) \cite{lewis_magnete_2014}& FM  & FM & 805 & 977 & 581 & 830 & 812 & 204.0 & 0.109 & 50.8 \\
CoPt \cite{wallace_magnetism_1971}& FM & FM & 813 & 651 & 642 & 664 & 659 & 134.3 & 0.089 & 52.5 \\
Fe$_3$O$_4$  \cite{levy_structure_2012}& FiM & FiM & 838 & 581 & 575 & 460 & 508 & 112.2 & 0.130 & 45.1 \\
FeNi$_3$ \cite{wallace_magnetism_1971}& FM  & FM & 863 & 1008 & 776 & 799 & 830 & 210.9 & 0.118 & 75.9 \\
Co$_2$MnSi \cite{brown_magnetization_2000} & FM & FM & 985 & 838 & 1141 & 668 & 765 & 181.9 & 0.120 & 112.4 \\
Fe (bcc) \cite{Ashcroft76}& FM  & FM & 1043 & 996 & 718 & 959 & 1037 & 208.3 & 0.095 & 68.4 \\
Co (fcc) \cite{Ashcroft76}& FM  & FM & 1388 & 792 & 594 & 1209 & 1429 & 163.9 & 0.055 & 47.4 \\

\end{tabular}
\end{ruledtabular}
\end{table*}

\begin{figure}

    \includegraphics[scale=0.35]{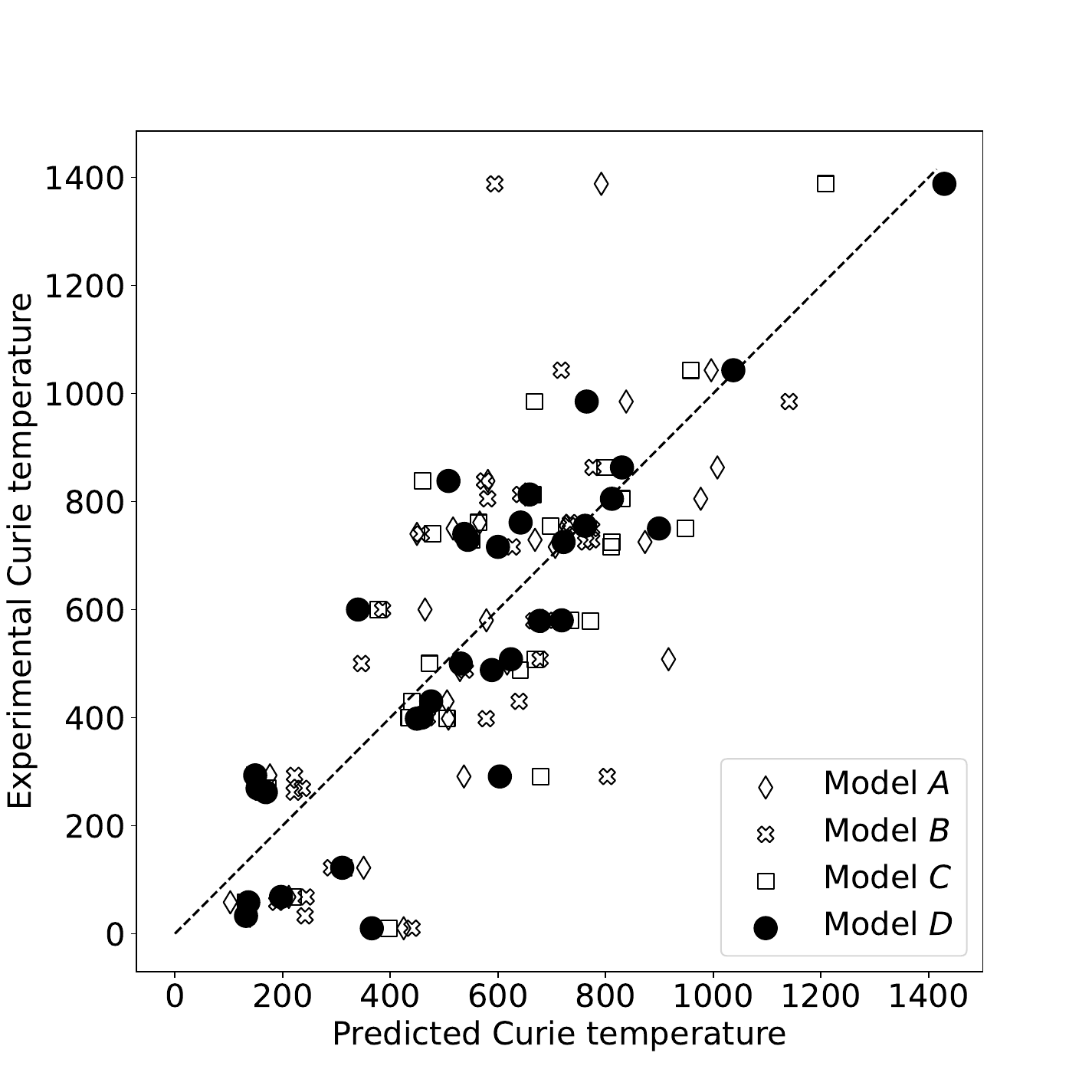}
    \caption{\label{fig:actual_predicted_all} Experimental versus predicted Curie temperatures based on the four models from Table \ref{tab:models}.}
    
\end{figure}

\begin{figure*}
 \subfloat[\label{abs_error}]{\includegraphics[scale=0.35]{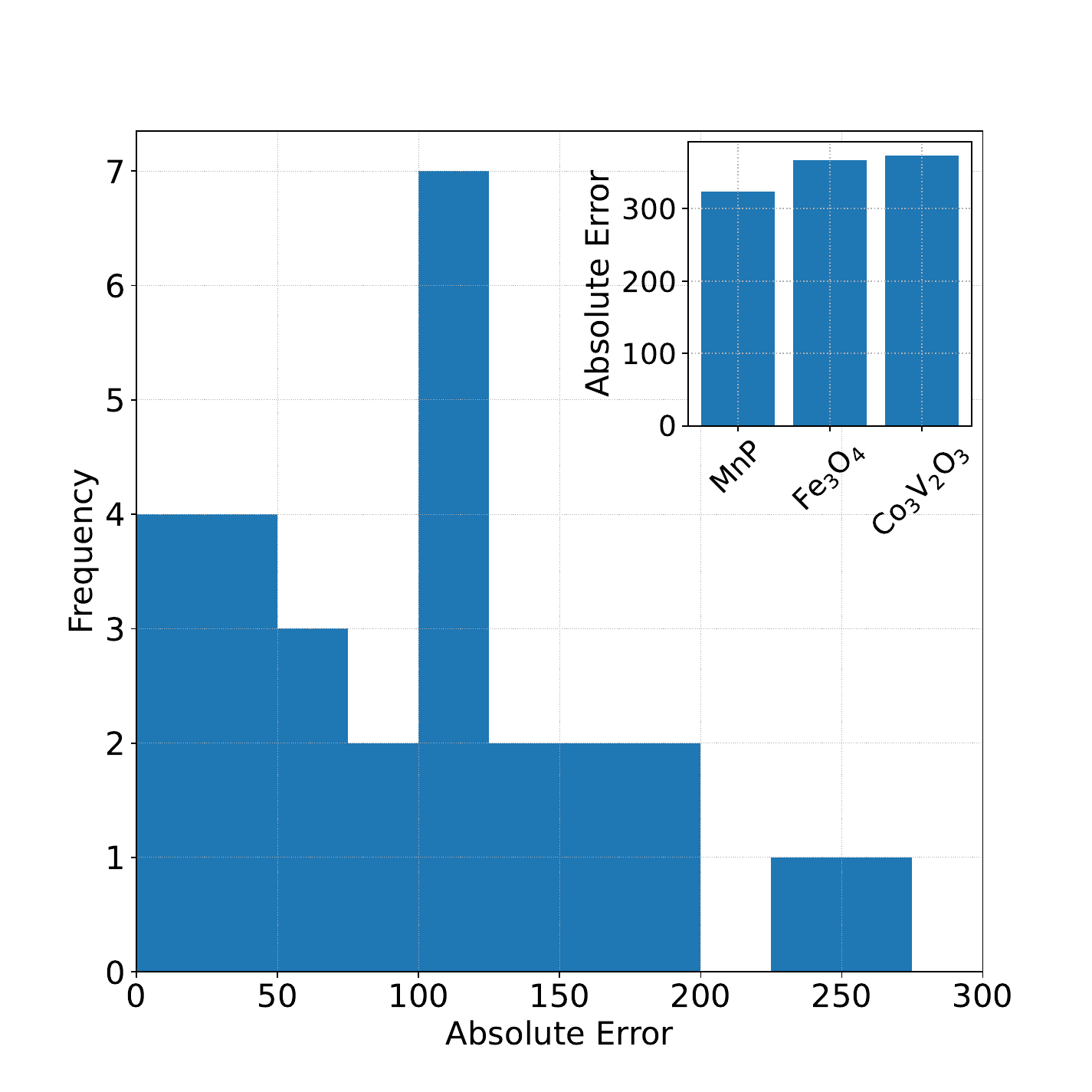}}
\subfloat[\label{relative_error}]{\includegraphics[scale=0.35]{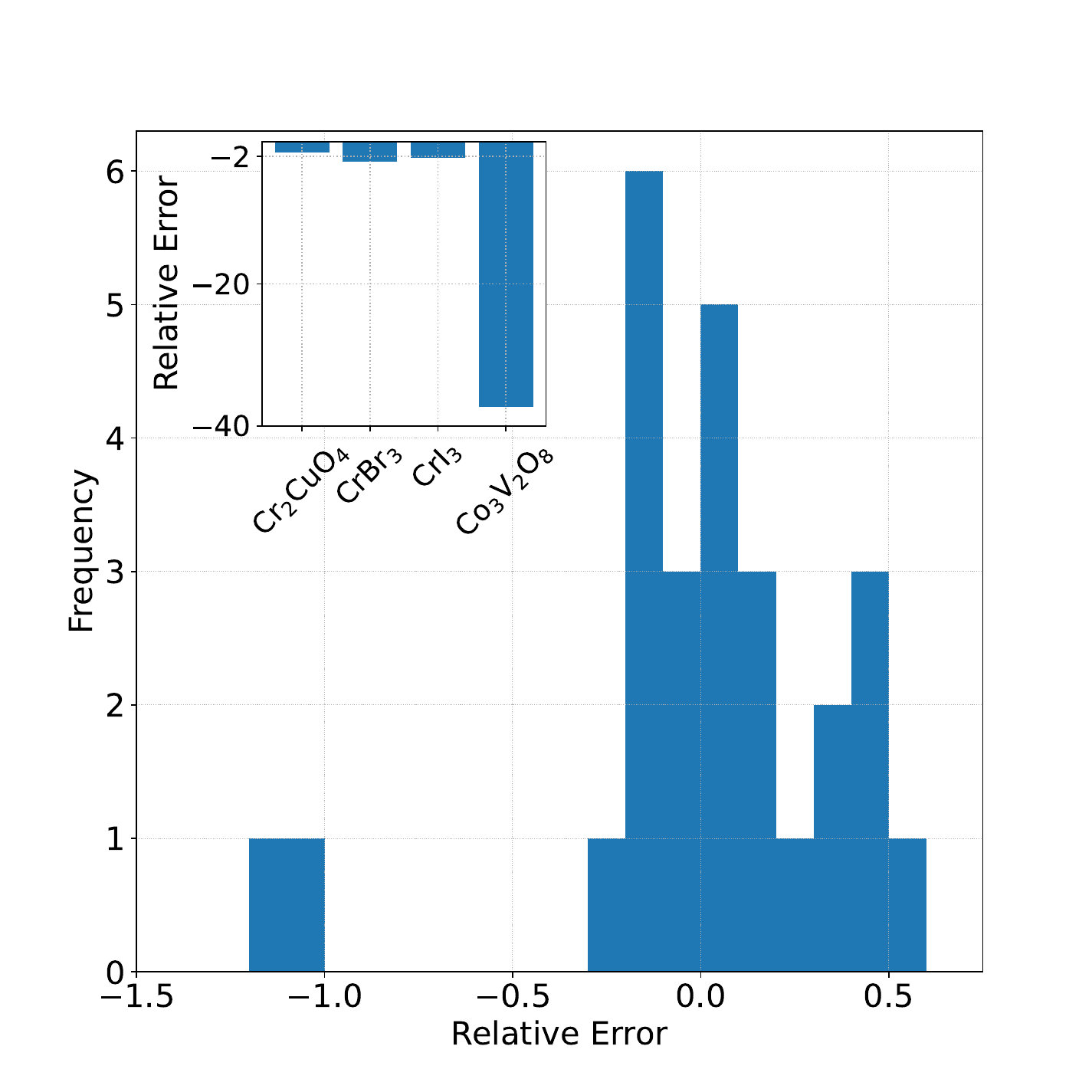}}

\caption{\label{fig:errors_S_NN} The absolute errors $\abs{T_\text{C}^\text{exp.} - T_\text{C}^\text{D}}$ and the relative errors $(T_\text{C}^\text{exp.} - T_\text{C}^\text{D})/T_\text{C}^\text{exp.}$ of model D. The inset in each subplot shows the systems with errors outside of the intervals on the $x$-axes.}
 \end{figure*}

\begin{table*}
\caption{\label{tab:models}The three initial models and their DFT input variables, final expressions, mean absolute errors, and $R^2$ values for the data set of materials discussed in Sections \ref{result}-\ref{discussion}. A perfect agreement would have $R^2=1$. The DFT input of models A and B are given per constrained atomic magnetic moment.}
\begin{ruledtabular}
\begin{tabular}{ccccc}
Model & DFT input &  Model (K) & Mean Absolute Error (K)& $R^2$ \\
\hline
A & $x = \Delta E$ [eV] & $ T_\text{C}^\text{A} = 4147x + 104$ & 160  & 0.64  \\
B & $x = B_\text{con}$ [eV] & $ T_\text{C}^\text{B} = 8126x + 198$ & 176   & 0.51\\
C & $x = \Delta E / S^\text{mag}$ [K] & $ T_\text{C}^\text{C} = 0.365x + 130 $& 157  & 0.67  \\
D &$x = \Delta E / S^\text{mag}$ [K] & $T_\text{C}^{D} = 0.85x \Big ( 1 - \frac{0.69}{NN^{0.14}}\Big ) + 124$ & 126 & 0.77 \\
\end{tabular}
\end{ruledtabular}
\end{table*}
\noindent
All properties needed for the four models described in section \ref{ref:model_devel} are calculated for 32 different systems, and results from the DFT calculations along with the predicted $T_\text{C}$ of the four models are presented in Table \ref{tab:dataset}. Figure \ref{fig:actual_predicted_all} shows the experimental $T_\text{C}$ against the predicted $T_\text{C}$ for each of the four models. In Table \ref{tab:models}, the expressions of the respective models are found along with their mean absolute error (MAE) and $R^2$-value. The $R^2$-value is given by
\begin{equation}
    R^2 = 1 - \frac{\sum_i [y^i - f(\pmb{x}^i)]^2}{\sum_i[y^i - \mu]^2}
\end{equation}
\noindent
where $y^i$ are target values, with mean $\mu$, and $f(\pmb{x}^i)$ are predicted values. The $R^2$ value and MAE are determined using a 3-fold cross validation scheme for the mean value of the respective errors. For model D, the resulting adjustment after optimization gives A $ = 0.69$ and B $ = 0.14$ in Equation (\ref{eq:adjustment_NN}). The final model D is given by,
\begin{equation}\label{eq:model_D}
    T_\text{C}^{\text{D}}= 0.85x \Bigg ( 1 - \frac{0.69}{NN^{0.14}}\Bigg ) + 124 \: \text{K},
\end{equation}
\noindent
where $x = \Delta E/S^\text{mag}$. The distribution of the absolute errors $\abs{T_\text{C}^\text{exp.} - T_\text{C}^\text{D}}$ and the relative errors $(T_\text{C}^\text{exp.} - T_\text{C}^\text{D})/T_\text{C}^\text{exp.}$ are shown in Figure \ref{fig:errors_S_NN} for model D.

\subsection{\label{ref:Fe-Co-alloy_result} Fe$_{1-x}$Co$_x$ Alloy}
Estimating the Curie temperature of various compositions in binary alloys is of great interest as it enables the selection and optimization of materials tailored to specific applications. Here, we apply model D of Equation (\ref{eq:model_D}) to the disordered Fe$_{1-x}$Co$_x$ alloys to investigate its effectiveness in making predictions for disordered binary alloys over compositions of different ratios of Fe and Co, for both the bcc and fcc structures. Figure \ref{fig:Fe-Co} shows predicted $T_\text{C}$ as a function of number of electrons per atom, going from pure bcc Fe to pure fcc Co. The experimental $T_\text{C}$ from Nishizawa and Ishida, and Rajeevan et al. are also included \cite{nishizawa_cofe_1984, rajeevan_structural_2022}. 
The experimental results from Refs. \cite{69Stu,75Nor} were compiled into a phase diagram of the magnetic transition in Fe$_{1-x}$Co$_x$ by Nishizawa and Ishida. The experimental point from Rajeevan et al. is for a bcc structure \cite{rajeevan_structural_2022}. For the Fe$_{1-x}$Co$_x$ alloy, the magnetic transition coincides with the bcc-fcc transition at approximately 31 at.\% Fe and 78 at. \% Fe in the Co- and Fe-rich regions, respectively \cite{nishizawa_cofe_1984}.
\section{Discussion \label{discussion}}
\subsection{Model Performance}

A general correlation of co-variation is found between the different input factors of the four different models and the experimental $T_\text{C}$. As expected, higher values correspond to higher experimental $T_\text{C}$, however, there is some significant noise. Mostly, model D performs better than the remaining three models, as can be seen in Figure \ref{fig:actual_predicted_all} and Table \ref{tab:models}. For some of the materials, the different models give vastly different predictions. Pure face-centered-cubic (fcc) Co is such a case. The fcc Co has relatively small magnetic moments (approximately 1 $\mu_B$) in the DLM simulations, and including the effect of magnetic entropy, as in models C and D, makes the estimations significantly better. A similar tendency is seen for FeNi and Cu$_2$MnIn where, in both cases, model A gives an estimate of $T_\text{C}$ that is 100 K too large, and model B more than 100 K too small. Models C and D estimate $T_\text{C}$ to be less than 50 K from the experimental value. There are some cases where model A gives a significantly better estimate, for example for FeB and MnP. However, the estimate for MnP is still quite poor when using either of the models. Model B gives a better estimate for hcp Gd and GdCd which could be due to the need for GGA+U to treat Gd correctly. When looking at the average strength of the constraining fields, the impact of not going beyond GGA in the calculations may not be as prominent since it is not dependent on the ground-state energy.  We can also see that in general, materials with very low $T_\text{C}$ are difficult for all of the models. Estimations for low $T_\text{C}$ materials are also difficult for ML models \cite{nelson_predicting_2019, belot_machine_2023}. The reason for these difficulties could be attributed to these systems having more complicated magnetic transitions to a larger extent than robust ferromagnets.

 As mentioned, model D performs significantly better than the other models in most cases. The adjustment based on the number of strong magnetic interactions, which incorporates structural information, clearly has a notable impact on the performance of the model. For model D, the $R^2$ value is improved to 0.77 and the MAE to 126 K. Figure \ref{fig:errors_S_NN} shows the absolute and relative errors of model D. More than 60 \% of the systems have an absolute error equal to or less than 126 K. As the MAE indicates, the concentration of the absolute errors is not towards zero, but rather around 126 K. The systems with the less accurate estimates have an absolute error of about 300 K. The relative errors are the largest for the systems with low $T_\text{C}$, where  Co$_3$V$_2$O$_8$ in particular stands out. This system has the highest absolute error. However, in general, the relative errors are less than 50 \% for a large majority of the systems, and approximately 60 \% of the systems have a relative error of less than or equal to 25 \%. For comparison, the best model of Nelson and Sanvito, which considers only chemical composition, has an MAE of approximately 50 K and $R^2$ value of 0.81\cite{nelson_predicting_2019}.
\subsection{\label{ref:Fe-Co-alloy_discussion} Fe$_{1-x}$Co$_x$ Alloy}
\begin{figure}

    \includegraphics[scale=0.34]{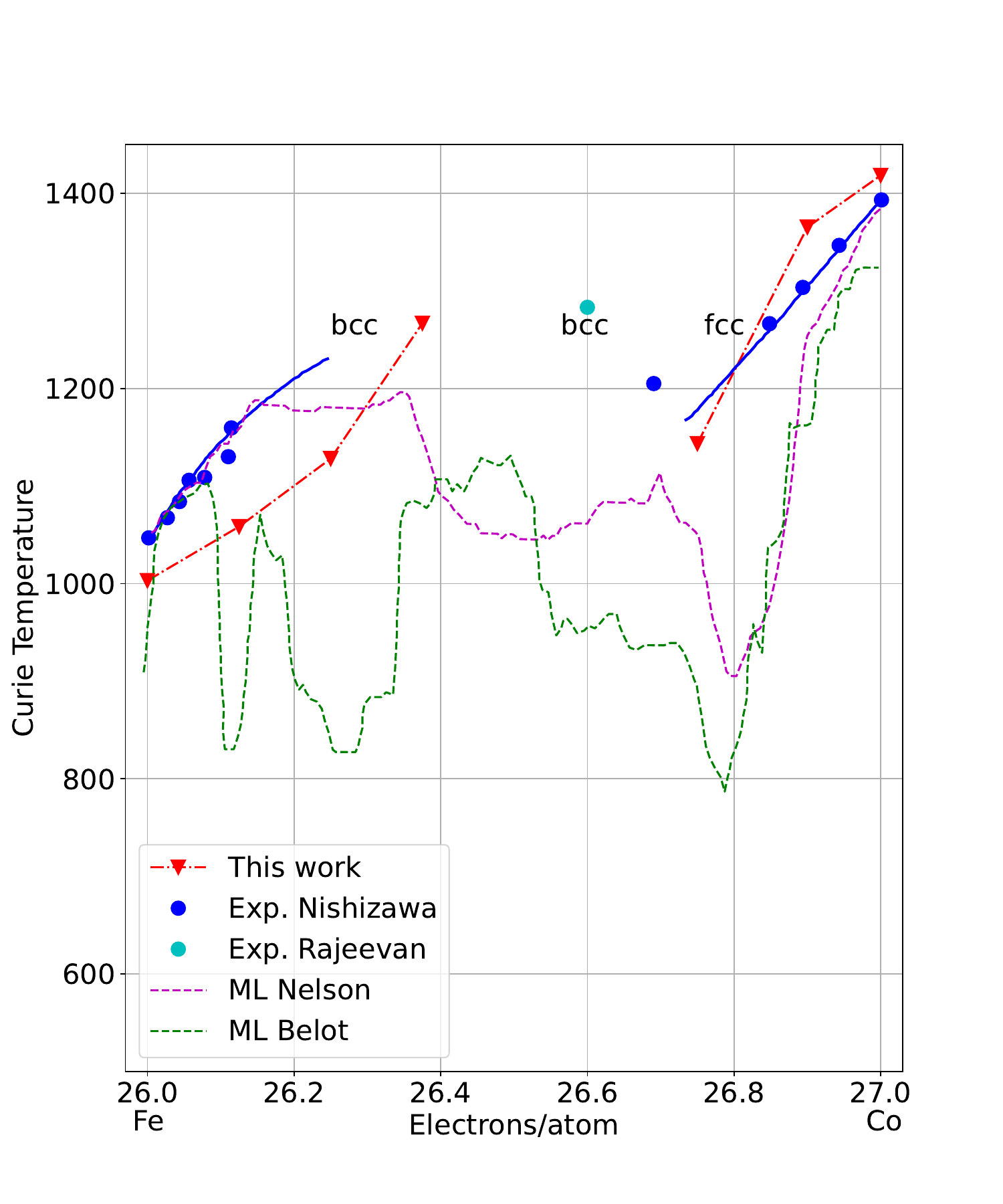}
    \caption{\label{fig:Fe-Co}Experimental and predicted Curie temperatures for the binary Fe$_{1-x}$Co$_x$ alloy as a function of electrons per atom. Estimates using our model D (red triangles, dashed line) for the disordered bcc structure for high Fe content (left) and the fcc structure for low Fe content (right). The experimental Curie temperatures compiled by Nishizawa and Ishida \cite{nishizawa_cofe_1984} (dark blue circles) with solid lines for their interpolation between points. The experimental point from Rajeevan et al. \cite{rajeevan_structural_2022} (turquoise circle) is for Fe$_{40}$Co$_{60}$. ML predictions (dashed blue and magneta lines) reproduced from Belot et al. \cite{belot_machine_2023} and Nelson and Sanvito \cite{nelson_predicting_2019}. The crystal structues are indicated in the plot (bcc or fcc).}
\end{figure}

 From Figure \ref{fig:Fe-Co}, we see that our model D captures the overarching trend well in both disordered bcc and fcc Fe$_{1-x}$Co$_x$ alloys. A clear correlation is shown where a higher Co content in the bcc structure gives a higher $T_\text{C}$ when moving towards the bcc-fcc phase boundary. Similarly, for the fcc structure, more Fe results in a lowering of $T_\text{C}$.  The results suggest the feasibility of obtaining a good estimate of the disordered binary alloy $T_\text{C}$ through our model, which could be used as a basis for making more complex calculations or experiments. For comparison with other methods for $T_\text{C}$ estimations, we have included the predictions of the same system by Nelson and Sanvito, and Belot et al. \cite{belot_machine_2023, nelson_predicting_2019} in Figure \ref{fig:Fe-Co}. In both of these cases, unphysical drifts of the predicted Curie temperatures are observed. As pointed out by Belot et al., these drifts occur between points of known experimental data included in the training data. This suggests a lack of transferability in the ML methods where the model does not capture the connections between different compositions of a binary alloy. In both studies, they attempted to include structural information into the descriptors, however this requires a larger feature space and leads to a smaller dataset due to the limited amount of structural data. As a result, the inclusion of structural information did not improve the model. The failure of ML methods in this case puts a critical perspective on the  good agreement with experiments reported when considering ordered structures. For disordered structures such as these, a method based on DFT calculations and straightforward physical assumptions, as used in the models of the present work, clearly has an advantage.

\subsection{\label{ref:comp_cost}Computational Cost}
Both the ground-state search and the DLM calculations are noncollinear magnetic calculations which are more computationally demanding than collinear magnetic calculations. Computations were performed using compute nodes with Intel Xeon Gold 6130 CPUs. The main part of the computational cost is the DLM step for almost all materials, which requires, on average, approximately 200 core hours; for the most costly calculations, up to 900 core hours are needed. One ground-state search takes on average 48 core hours which we may compare with the ferromagnetic collinear calculation which takes, on average, 4 core hours for the same structure. The noncollinear magnetic ground-state search is therefore, on average, 12 times more time-consuming than a straightforward collinear ferromagnetic calculation. For a few materials, it is even several hundred times more time-consuming. This is a significantly larger computational cost, but as we have seen, noncollinear spins are necessary to ensure finding the DFT ground state in some cases.

\section{Conclusions}
In this work we have investigated how the Curie temperature can be estimated based on the energy difference between the paramagnetic state and the ground state obtained using first-principles DFT calculations. We find that by introducing the effect of the magnetic entropy in the paramagnetic state, effectively also accounting for the magnetic moment sizes of different systems, we manage to predict $T_\text{C}$ with an $R^2$ value of 0.67 and a mean absolute error of 157 K. By also including the number of nearest magnetic neighbors, influencing the effect of magnetic SRO on $T_\text{C}$, we can further improve the model to get an $R^2$ value of 0.77 and a mean absolute error of 126 K. 

In this study, we have also explored the ability of our approach to describe the change in Curie temperature in a disordered binary alloy as a function of the composition. We find that our model D estimates the general trend of $T_\text{C}$ well across the range of compositions for the Fe$_{1-x}$Co$_x$ disordered alloy. The ability to predict such trends can be used, e.g., to predict the tuning of the Curie temperature of an alloy for specific applications. It may be possible for an ML model trained for $T_\text{C}$ predictions for the materials in this work to give a lower MAE and higher $R^2$-value, however, the substantially improved estimations for various compositions of the Fe$_{1-x}$Co$_x$ alloy underscore the necessity of incorporating physics-based insights when predicting the Curie temperature of a magnetic material. 

Our approach can be used as an early screening step in high-throughput workflows. The model can efficiently identify candidates for further exploration which are then investigated using computationally heavy calculations or experiments. Furthermore, the method can be combined with ML models to improve their performance in e.g., a delta learning approach where one trains a model to predict the differences between one of our models and the true value. This approach may help the physical insights of the DFT-based model to be integrated in the ML models and act as a substitute for experimental training data where it is lacking. This integration can potentially lead to enhanced predictive capabilities in an ML-based screening for new magnetic materials.

\begin{acknowledgments}
The computations were enabled by resources provided by the National Academic Infrastructure for Supercomputing in Sweden (NAISS) partially funded by the Swedish Research Council through grant agreement no. 2022-06725. B.A. acknowledges financial support from the Swedish Research Council (VR) through Grant No. 2019-05403, and 2023-05194 and from the Swedish Government Strategic Research Area in Materials Science on Functional Materials at Link\"oping University (Faculty Grant SFOMatLiU No. 2009-00971). R.A. acknowledges financial
support from the Swedish Research Council (VR)
Grant No. 2020-05402 and the Swedish e-Science Research
Centre (SeRC).
\end{acknowledgments}

\bibliographystyle{apsrev4-2}
\end{document}